\documentclass[pss]{wiley2sp} 

\usepackage{amsmath}
\usepackage{siunitx}
\usepackage{color}

\DeclareMathSymbol{\Omega}{\mathalpha}{letters}{"0A}
\DeclareMathSymbol{\varOmega}{\mathalpha}{operators}{"0A}

\begin{document}

\title{Quantized electron transport through graphene nanoconstrictions}

\titlerunning{Quantized transport through graphene nanoconstrictions}

\author{%
  V.\ Cleric\`{o}\textsuperscript{\textsf{\bfseries 1}},
  J.\ A.\ Delgado-Notario\textsuperscript{\textsf{\bfseries 1}},
  M.\ Saiz-Bret\'{\i}n\textsuperscript{\textsf{\bfseries 2}},
  C.\ H.\ Fuentevilla\textsuperscript{\textsf{\bfseries 1}},
  A.\ V.\ Malyshev\textsuperscript{\textsf{\bfseries 2}},
  J.\ D.\ Lejarreta\textsuperscript{\textsf{\bfseries 1}},
  E.\ Diez\textsuperscript{\textsf{\bfseries 1}},
  F.\ Dom\'{\i}nguez-Adame\textsuperscript{\Ast,\textsf{\bfseries 2}}
}

\authorrunning{Cleric\`{o} \emph{et al.}}

\institute{%
  \textsuperscript{1}\,NANOLAB, Departamento de F\'{\i}sica Fundamental, Universidad de Salamanca, E-37008 Salamanca, Spain\\
  \textsuperscript{2}\,GISC, Departamento de F\'{\i}sica de Materiales, Universidad Complutense, E-28040 Madrid, Spain}

\received{XXXX, revised XXXX, accepted XXXX} 

\published{XXXX} 


\abstract{%
\abstcol{%
We study the quantization of Dirac fermions in lithographically defined graphene nanoconstrictions. We observe quantized conductance in single nanoconstrictions fabricated on top of a thin hexamethyldisilazane layer over a Si/SiO$_2$ wafer. This nanofabrication method allows us to obtain well defined edges in the nanoconstrictions, thus reducing the effects of edge roughness on the
}{%
conductance. We prove the occurrence of ballistic transport and identify several size quantization plateaus in the conductance at low temperature. Experimental data and numerical simulations show good agreement, demonstrating that the smoothing of the plateaus is not related to edge roughness but to quantum interference effects.
}}

\maketitle

\section{Introduction}

Electronic excitations in the vicinity of the Fermi level of a number of advanced materials resemble massless Dirac fermions. The band structure close to the Fermi level of these materials, referred to as Dirac materials~\cite{Wehling14}, is characterized by the well-known linear energy-momentum relation of relativistic massless particles, hence the name of Dirac cones (see Refs.~\cite{Wehling14,Diaz-Fernandez17a,Diaz-Fernandez17b} and references therein). Dirac materials include a plethora of different systems such as $d$-wave superconductors, graphene, and topological insulators, to name a few. From the standpoint of applications, Dirac materials are envisioned to be of outstanding importance due to their universal behavior and the robustness of their properties~\cite{Wehling14}.

Among Dirac materials, graphene stands out because it possesses unique electronic, thermal and mechanical properties~\cite{Novoselov04,Castro09}. Since graphene is envisioned as a material of choice for a variety of applications in future electronics, a great effort is being made to understand electron transport properties in nanostructures based on graphene. Narrow graphene stripes, known as graphene nanoribbons~(GNRs), are the most basic building blocks for other graphene nanodevices~\cite{Bischoff14} such as nanorings~\cite{Luo09,Wu10,Munarriz11,Munarriz12,Benjamin14,Yannouleas15a,Jang15} and superlattices~\cite{Sevincli08,Munarriz13,Dubey13,Diaz14,Ban15,Lee16,Tian16,Zhang16} (see Ref.~\cite{Yagmurcukardes16} for a recent review on nanoribbons of two-dimensional materials). GNRs behave as single-channel room-temperature ballistic electrical conductors on a length scale greater than ten microns~\cite{Baringhaus14}. Therefore, they are good candidates to exploit quantum effects such as the Fano effect~\cite{Gong13,Xu13,SaizBretin15a,SaizBretin15b,Petrovic15,Gehring}, resonant tunneling~\cite{Britnell13,Gaskell15,GuerreroBecerra16} and quantum size effects~\cite{Huang14,Zhu17}, even at room temperature. GNRs are commonly referred to as graphene nanoconstrictions~(GNCs) or quantum point contacts when their length is close to their width.

A number of different fabrication methods make it possible to achieve GNCs with high crystalline quality. On a microscopic level, however, imperfections such as unintentional adsorbed atoms, charged impurities in the substrate, vacancies and edge disorder, will strongly depend on the method used to fabricate the samples. These defects may alter the electron transport properties to some extent. On one hand, electron transport in graphene is known to be less influenced by metal adatoms than other honeycomb-lattice materials like silicene~\cite{Lin12,Ersana14,Du14,Nunez17}. On the other hand, charged impurities of the substrate would provide some additional smooth electrostatic potential and can hardly deteriorate the electron transmission through the device. The impact of edge disorder on the transport properties, however, is expected to be stronger, especially for small devices.

The aim of this paper is to investigate electron transport properties and size quantization effects in GNCs. We have developed a reliable fabrication method that enables an optimal electron beam lithography~(EBL) processing of the GNC while retaining high carrier mobility. Low temperature conductance measurements performed on narrow GNCs present clear signatures of coherent transport and size quantization effects. However, conductance plateaus are smooth and not so well defined as in long GNRs. To understand the observed smoothing of the conductance plateaus we also carried out numerical simulations within the nearest neighbor tight-binding approach for $p_z$ electrons of C atoms in graphene~\cite{Munarriz11,Munarriz12,SaizBretin15a,SaizBretin15b}. Simulations of GNCs samples without edge roughness do not show abrupt plateaus which, however, are revealed in long GNRs. The good agreement between theory and experiment leads us to the conclusion that quantum interference is responsible for the smooth plateaus observed in the conductance while edge disorder does not play any relevant role. 

\section{Experimental methods}

We have fabricated GNCs by EBL of mechanical exfoliated graphene flakes transfered to Si/SiO$_2$ wafers treated before and after the exfoliation with hexamethyldisilazane~(HMDS), following a si\-mi\-lar procedure to the one used by Caridad \emph{et al.}~\cite{Caridad}. It is expected that graphene deposited on a hydrophobic subs\-trate has better transport pro\-per\-ties due to the HMDS immersion. In particular, these samples show higher carrier mobility and Dirac peaks closer to zero due to the low interaction of graphene with the substrate~\cite{Lafkioti,Chouwdhury}.

\begin{figure}[ht]
\centering{\includegraphics[width=0.8\columnwidth]{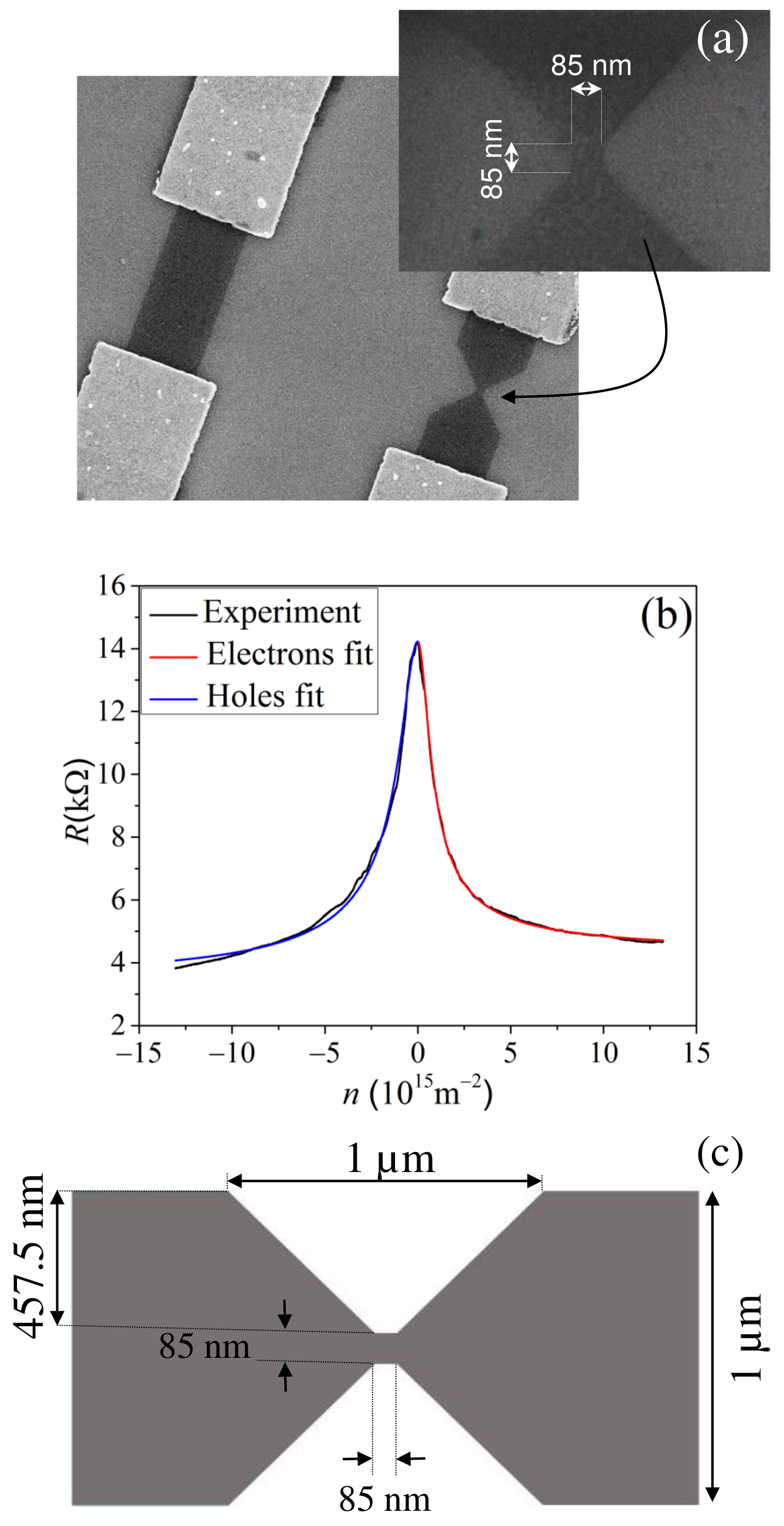}}
\caption{(a)~SEM image of a $\SI{1}{\micro\metre}\times\SI{2}{\micro\metre}$ bar (left) and $\SI{85}{\nano\metre}\times\SI{85}{\nano\metre}$ GNC (right). Scale bar is given by the width of the stripe ($\SI{1}{\micro\metre}$). The inset shows an enlarged view of the GNC. (b)~Resistance $R$ as function of the carrier density in the bar. Red and blue lines are the fitting of the data with Eq.~(\ref{eq:02}). (c)~Geometrical factors used to obtain the number of squares~$N_\mathrm{sq}$ in Eq.~(\ref{eq:02}).}
\label{fig1}
\end{figure}

The substrate is a $n$-doped Si wafer with a $\SI{290}{\nano\metre}$ layer of SiO$_{2}$. The wafer was dipped in a $1\!:\!1$ solution of HMDS and acetone for at least  $18$ hours before transferring the graphene flake. Afterwards, graphene is deposited on the  wafer and identified by means of optical microscopy and Raman spectroscopy. After Raman characterization, a monolayer graphene flake is chosen and ohmic contacts are defined by EBL and subsequent evaporation of Au and Ti. A second step of EBL defining a PMMA mask for the  etching process combining inductively coupled plasma and reactive ion etching is required. We have used an  O$_{2}$ and Ar atmosphere to define the $\SI{85}{\nano\metre}$ GNC and a $\SI{1}{\micro\metre}\times\SI{2}{\micro\metre}$ bar. The device is then suitable for transport measurements after a se\-cond immersion in HDMS for at least $24$~hours. In Figure~\ref{fig1}(a) we show a SEM image of our  devices, namely a $\SI{1}{\micro\metre}\times\SI{2}{\micro\metre}$ graphene bar and a $\SI{85}{\nano\metre}\times\SI{85}{\nano\metre}$ GNC. 

Similar GNCs were studied based on he\-te\-ros\-tructures of graphene and hexagonal boron nitride (hBN) by Terr\'{e}s \emph{et al.}~\cite{Terres}. In spite of the high electronic mobility, they restricted their study to GNCs larger than $\SI{200}{\nano\metre}$ due to several issues in the etching process of the he\-te\-ros\-tructure. On one hand, the  dielectric nature of hBN is not suitable for EBL masking. On the other hand, hBN requires the use of SF$_6$ atmosphere that increases the edge roughness during the etching process. Furthermore, the thickness of the heterostructure, usually larger than $\SI{20}{\nano\metre}$, demands aspect ratios much larger than in bare graphene monolayer flakes. In fact, the fabrication of GNCs with EBL on these samples is limited to rather larger sizes (minimum size $\SI{230}{\nano\metre}$~\cite{Terres}). 

Recently, another type of GNC in encapsulated graphene was studied for sizes less than $\SI{200}{\nano\metre}$, yet edge roughness was not negligible~\cite{Kumar}. Previous studies by Tombros et al. considered small GNCs in suspended graphene with very high mobility ~\cite{Niko}, although the channel was not well defined due to the fabrication technique used (high DC current during annealing to create a constriction without a control on the final width).   

Some results have been reported in GNCs of graphene over wafers of Si/SiO$_2$. Although these GNCs have low edge roughness and good control of the size, the electron mobility was found to be very low due to the direct contact of graphene and the oxide~\cite{Ozyilmaz,Lu,Gehring,Han} that prevents the observation of size quantization effects. For all these reasons, we believe that graphene on HMDS represents a good compromise between high mobility graphene and well defined constrictions with different geometries.

\section{Electron transport results}

We present here two-terminal transport measurements for GNCs fabricated on a graphene flake using the HDMS treatment described above. The results are representative and reproducible in similar devices with the same width and electron mobility. All transport measurements were taken at $\SI{5}{\kelvin}$. In Figure~\ref{fig1}(b) we show the resistance of the bar [device shown in the left side of Figure~\ref{fig1}(a)] as a function of the carrier density $n$ created by a back-gate voltage $V_\mathrm{bg}$. This carrier density can be estimated from  
\begin{equation}
n=\frac{C_\mathrm{ox}}{e}\,(V_\mathrm{bg}-V_\mathrm{CNP})=\frac{C_\mathrm{ox}}{e}\,V_\mathrm{bg}^{*}
\label{eq:01}
\end{equation}
where $C_\mathrm{ox}$ is the SiO$_{2}$ capacitance per area, $e$ the elementary charge and $V_\mathrm{CNP}$ the voltage in the Dirac point (less than $\SI{3}{\volt}$ in this device). We define also a normalized gate voltage as $V_\mathrm{bg}^{*} \equiv V_\mathrm{bg}-V_\mathrm{CNP}$. We can estimate the mobility, residual doping, and contact resistance by fitting our data shown in Figure~\ref{fig1}(b) to the following expression~\cite{Kim,Gammelgaard}
\begin{equation}
R=R_{c}+\frac{N_\mathrm{sq}}{e\mu\sqrt{n_0^2+n^{2}}}\ ,
\label{eq:02}
\end{equation}
where $R_{c}$ is the contact resistance, $N_\mathrm{sq}$ is the number of squares in graphene channel, that is, the length over width ratio, $n_{0}$ is the residual doping, $e$ is the ele\-men\-ta\-ry charge and $\mu$ is the field effect mobility. The values obtained for the bar with $N_\mathrm{sq}=L/W$ when $L=\SI{2}{\micro\meter}$ and $W=\SI{1}{\micro\meter}$ yield a contact resistance of $R_{c}=\SI{3.2}{\kilo\ohm}$ and a residual doping of $n_{0}=\SI{5.5e10}{\per\square\centi\meter}$. On the other hand, the mobility for the electrons is $\mu_{n}= \SI{2.0e4}{\square\centi\meter\per\volt\per\second}$ whereas the hole mobility is $\mu_{p}=\SI{1.3e4}{\square\centi\meter\per\volt\per\second}$. These values are comparable to those obtained in similar samples of graphene subjected to HMDS treatment~\cite{Hansel}, being higher than standard graphene samples on SiO$_{2}$. The value of $N_\mathrm{sq}$ in Eq.~(\ref{eq:02}) is obtained from
\begin{equation}
N_\mathrm{sq}=\frac{L^{2}}{WL-(b_1+b_2)h}\ ,
\label{eq:03}
\end{equation}
where $h=\SI{457.5}{\nano\meter}$, $b_1=\SI{1}{\micro\meter}$ and $b_2=\SI{85}{\nano\meter}$ are the geometrical parameters of the GNC shown in Figure~\ref{fig1}. The number of squares $N_\mathrm{sq}$ obtained from (\ref{eq:03}) is $2.66$, close to the previously estimated value $N_\mathrm{sq}=L/W=2$.

\begin{figure}[ht]
 \centering\includegraphics[width=0.7\columnwidth]{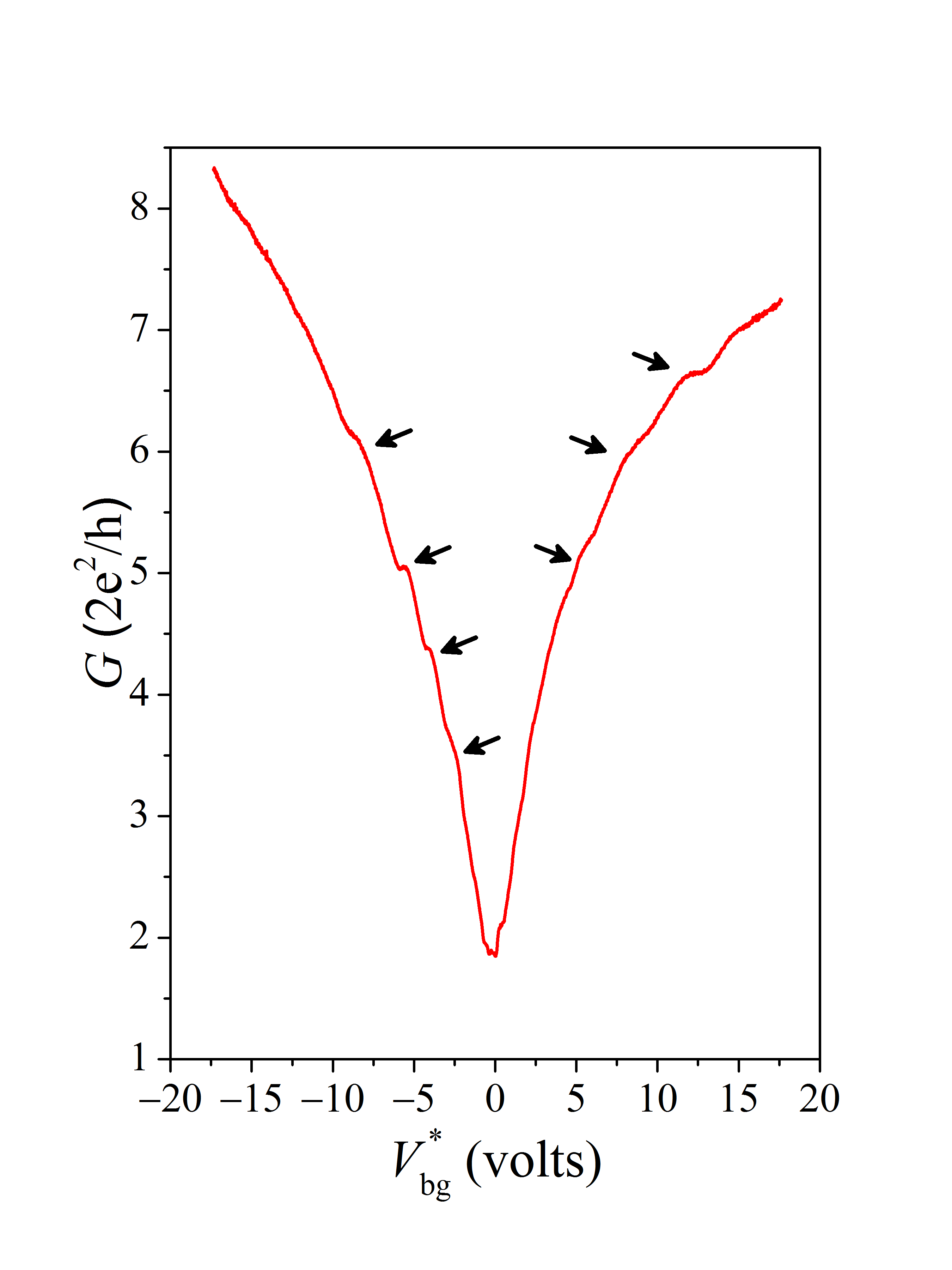}
\caption{Conductance versus the shifted back-gate voltage $V_\mathrm{bg}^{*}=V_\mathrm{bg}-V_\mathrm{CNP}$ for the GNC shown in Figure~\ref{fig1}. Black arrows indicate the observed \emph{kinks\/} at some particular values of the back-gate voltage.}
\label{fig2}
\end{figure}

The two-terminal conductance $G=N_\mathrm{sq}/R$ as a function of the shifted back-gate voltage $V_\mathrm{bg}^{*}$ is shown in Figure~\ref{fig2}. The curve shows a well-defined scaling $G\propto |V_\mathrm{bg}^{*}|^{1/2}$. This scaling is regarded as an indication of both ballistic transport regime and homogeneity of the graphene flake. In particular, this trend is very similar to the two-terminal transport measurements reported by Tombros \emph{et al.}~\cite{Niko} in high-mobility suspended  GNCs. Black arrows in Figure~\ref{fig2} highlight a set of reproducible \emph{kinks\/} in the conductance curve. These kinks are well reproduced in several cool-downs and even with measurements taken after several months and after additional steps of HDMS cleaning. The separation between two consecutive kinks is around $2e^{2}/h$, indicating quantized conductance through the single GNC. According to our simulations, the occurrence of kinks instead of sharp quantization steps can be explained solely by geometrical effects even in the presence of perfect edges. Therefore, there is no need to rely on strong scattering at rough edges, as claimed by Terres \emph{et al}.~\cite{Terres}.

\begin{figure}[ht]
 \centering{\includegraphics[width=0.7\columnwidth]{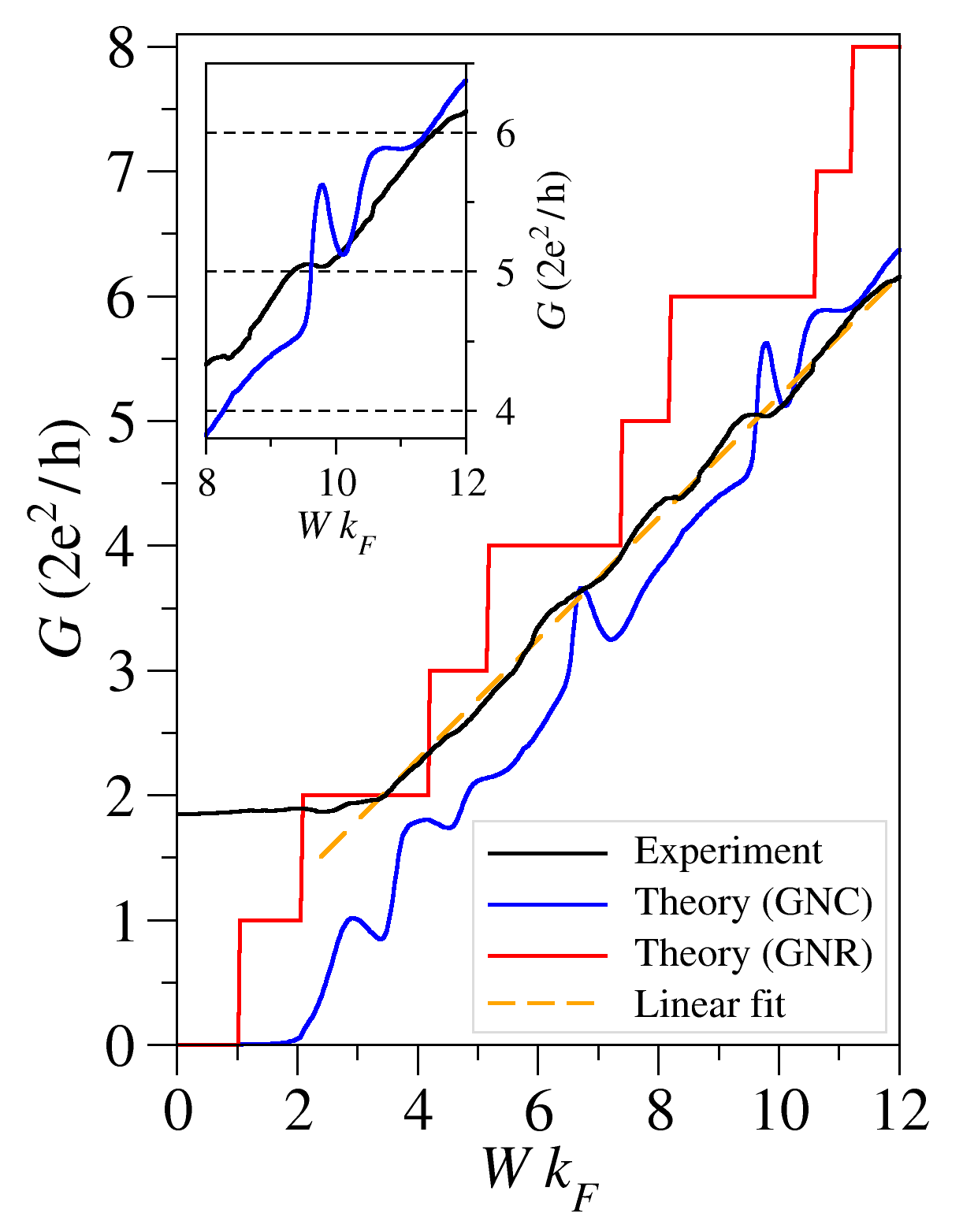}}
\caption{Conductance in units of $2e^2/h$ as a function of the dimensionless magnitude of $Wk_{F}$. The black solid line corresponds to the measurement at low temperature, showing the expected linear trend. The dashed line presents a linear fit to the experimental data. Numerical simulations corresponding to a GNC and a very long GNR of the same width $W=\SI{85}{\nano\meter}$ are shown in blue and red solid lines, respectively. The inset shows an enlarged view of the conductance at high electron density.} 
\label{fig3}
\end{figure}

Our claim that the conductance kinks are signatures of size quantization is supported by numerical simulations, which reproduce nicely the experimental results. Simulations were based on the quantum transmission boundary method~\cite{Lent90,Ting92} combined with the effective transfer matrix approach~\cite{Schelter10}, within the tight-binding approach for the $p_z$ electrons of graphene~\cite{Munarriz11,Munarriz12,SaizBretin15a,SaizBretin15b} with the boundary conditions that require that the wave function vanishes on fictitious sites outside the GNR (see Refs.~\cite{Wurm09,Munarriz14} for further details). We performed the vast majority of computer simulations of GNRs with armchair edges. GNRs with zig-zag edges were also considered and similar results were found, therefore we restrict ourselves to armchair edges hereafter.In Figure~\ref{fig3} we plot the conductance as a function of the dimensionless magnitude $Wk_{F}$, where $W$ is the width of the GNC and $k_{F}=\sqrt{\pi n}$ is the Fermi wave number. We neglect a small additional contribution due to the localized trap states that would slightly modify the relation between $n$ and $k_{F}$. The blue solid line shows the numerically simulated conductance of a GNC with straight edges whereas the black solid line corresponds to the measured conductance. Results of Figure~\ref{fig3} are similar to those by Yannouleas \emph{et al.}~\cite{Yannouleas15a,Yannouleas15b} who found spikes in nanoconstrictions conductance and attributed them to the longitudinal quantization due to the finite length of the GNC.For comparison, the red solid lines displays the conductance of a very long GNR when $W=\SI{85}{\nano\meter}$. In the absence of constriction, the conductance at low temperature presents well-defined steps of height $2e^2/h$ due to the quantization of the transverse momentum. A similar trend reported by Haanappel and van der Merel in a two-dimensional electron gas based on a GaAs heterostructure~\cite{Haanappel89}. It is worthwhile to make a comparison between electrostatically defined point contacts in GNRs and our samples where the GNC is defined using nanolithography. For instance, in narrow point contacts defined by electrostatic gating the conductance is characterized by abrupt changes and exhibits Fano resonances~\cite{Petrovic15}. All these features are absent in nanolithographycally fabricated GNCs.

Figure~\ref{fig3} shows that the experimental conductance achieves an almost constant value when $Wk_{F} <  4$. On the contrary, the  theoretical conductance displays a gap for $Wk_{F}<  2$. The non-zero minimum value of the experimental conductance is attributed to the residual doping in the graphene flakes, which prevents from achieving a zero carrier density experimentally. Accordingly, we cannot reach values of  $k_F$ below a minimum cut-off arising from the residual doping density $n_0$. Since $k_{F}=\sqrt{\pi n}$, this cut-off value can be obtained straightforwardly as $k^{\text{cut-off}}_F = \sqrt{\pi n_0} \sim  4.2\times\SI{e5}{\per\centi\meter}$ with $n_0 =\SI{5.5e10}{\per\square\centi\meter}$ being the residual doping obtained above from the fit of the data shown in Figure~\ref{fig1}(b). In terms of the dimensionless parameter we get $Wk^{\text{cut-off}}_F \sim  4$, in excellent agreement with the experimental data presented in Figure~\ref{fig3}. 

For values of the Fermi wave number larger than $k^{\text{cut-off}}_{F}$, the experimental and simulated conductance curves present a linear dependence on $k_{F}$, as expected for a ballistic regime. The slopes of both lines are quite close, and the small differences arise from imperfections, i.e. smoother edges at the corners of the GNC in real samples. We can consider the conductance within the Landauer approach~\cite{Terres}
\begin{equation}
G=\left(\frac{2 \, e^{2}}{h}\right)\frac{2Wk_{F}}{\pi} \, c_{0}
\label{eq:04}
\end{equation}
for a GNC of width $W$, where $c_{0}$ is a pa\-ra\-me\-ter between $0$ and $1$ related to edge roughness. For straight edges this parameter becomes $c_{0}=1$ while it is smaller when edge roughness is not negligible. We can estimate the value of $c_{0}$ from the linear fit of the experimental conductance shown in Figure~\ref{fig3} (dashed line). In our GNC with $W=\SI{85}{\nano\meter}$ we found $c_{0}=0.74$, suggesting low edge roughness of the sample. For comparison, previous results by Terres \emph{et al.}~\cite{Terres} in hBN/graphene/hBN GNCs reported values about $c_{0}=0.56$ in much wider GNCs ($W=\SI{230}{\nano\meter}$). Therefore, we claim that our smaller GNC is less affected by edge disorder on electron transport as compared to the larger ones based on encapsulated graphene~\cite{Terres}. This result may seem counterintuitive and can be explained as follows. The fabrication of GNC on HMDS allows for a better definition of the edges while overcoming some of the limitations that occur in processing encapsulated graphene. Also, we cannot rule out a higher concentration of vacancies \cite{Petrovic16} in the GNCs fabricated on heterostructures of graphene and hBN due to the transfer technique reported in Ref.~\cite{Terres}. This unavoidable vacancy disorder can smear out quantized conductance steps.

Finally, in the inset of Figure~\ref{fig3} we show an enlarged view of the kinks in the conductance curve, revealed now  as clear quantization steps for some specific values of $Wk_{F}$. Although  we do not observe all the features expected for a perfect GNC, the qualitative agreement of experimental and simulated conductances allow us to unambiguously identify these quantization steps as size quantization signatures. 

\section{Conclusion}

We studied theoretically and experimentally the electron transport properties of a $\SI{85}{\nano\meter}$ GNC using a HMDS treatment to enhance carrier mobility. A remarkable result is that the use of this HMDS treatment for graphene nanoconstrictions re\-pre\-sents a good compromise between well defined constriction and high mobility. In comparison to recent works on constrictions of encapsulated graphene into hBN, we find a much lower roughness of the edge and consequent lower scattering effects, signs of a better quality and good edge definition of the GNC. Using low-temperature transport measurements, we have demonstrated that the conductance is ballistic. It also manifests clear steps resulting from size quantization, which is in good agreement with our numerical simulations.

\begin{acknowledgement}

The authors are indebted to A. D\'{\i}az-Fern\'{a}ndez and J. M. Caridad for enlightened discussions and for a critical reading of the manuscript. This research has been supported by MINECO (Grants MAT2013-46308 and MAT2016-75955) and Junta de Castilla y Le\'{o}n (Grant SA045U16).

\end{acknowledgement}

\bibliographystyle{pss}

\bibliography{references}

\end{document}